\documentclass[%
preprint,
]{revtex4-2}

\usepackage{chemformula} 
\usepackage[T1]{fontenc} 

\usepackage{amsfonts}

\usepackage{mathptmx}
\usepackage[utf8]{inputenc}
\usepackage[hidelinks]{hyperref}
\usepackage{braket}
\usepackage{booktabs}
\usepackage{array}

\def\C{{\mathbb C}}
\def\R{{\mathbb R}}
\def\N{{\mathbb N}}
\def\id{{\rm Id}}

\newcommand{\bigo}{ \mathcal{O} }
\newcommand{\re}[1]{ \text{Re} \left( #1 \right)}

\newcommand{\sign}[1]{ \text{sign} \left( #1 \right)}

\newcommand{\adi}{\text{adi}}
\newcommand{\dia}{\text{dia}}
\newcommand{\el}{\text{el}}
\newcommand{\nuc}{\text{nuc}}
\newcommand{\out}{\text{out}}
\newcommand{\stay}{\text{stay}}

\DeclareMathOperator{\diag}{\text{diag}}

\begin{document}
\author{Leonardo Araujo}
\email{leonardo.araujo@tum.de}
\author{Caroline Lasser}
\email{classer@ma.tum.de}
\affiliation{Department of Mathematics, TUM School of Computation, Information
	and Technology, Technische Universit\"{a}t München, Boltzmannstr. 3, 85748 Garching bei M\"{u}nchen}
\author{Burkhard Schmidt}
\email{burkhard.schmidt@wias-berlin.de}
\affiliation{Weierstrass Institute for Applied Analysis and Stochastics, Mohrenstr. 39, 10117 Berlin}

\title[FSSH-2]
{FSSH-2: Fewest Switches Surface Hopping with robust switching probability}
	


%
%



\begin{abstract}
	This study introduces the FSSH-2 scheme, a redefined and numerically stable adiabatic Fewest Switches Surface Hopping (FSSH) method for mixed quantum-classical dynamics. 
	It reformulates the standard FSSH hopping probability without non-adiabatic coupling vectors and allows for numerical time integration with larger step sizes.
	The advantages of FSSH-2 are demonstrated by numerical experiments for five different model systems in one and two spatial dimensions with up to three electronic states.
\end{abstract}

\maketitle


\section{Introduction}

The widely--used Fewest Switches Surface Hopping (FSSH) method is a cost--effective approach for the numerical simulation of non--adiabatic effects in quantum--classical dynamics.
Even though originally developed more than 30 years ago~\cite{Tully1990}, this concept is still widely used, especially in the context of photochemistry and excited-state molecular dynamics ~\cite{Crespo2018,Ibele2020,Jain2022,Freixas2021}. 
The FSSH scheme adopts a mixed quantum--classical paradigm where typically nuclei are treated classically and electrons quantum-mechanically. 
The method's principal efficiency stems from its use of classical trajectories moving on adiabatic potential energy surfaces with the possibility of stochastic 'hopping' between states according to calculated probabilities. 
Because of this conceptual simplicity and the straightforward implementation, FSSH is still among the most commonly used
methods for non--adiabatic molecular dynamics.

Despite its efficiency, FSSH has the drawback that it relies on the notorious Non--Adiabatic Couplings (NACs). 
While there are methods to compute NACs accurately and efficiently in regions of large gaps between neighboring potential energy surfaces \cite{Hammes1994,Jain2016,Jain2022}, they may become increasingly large near energy crossings, if not divergent.
For trajectories approaching genuine or avoided crossings, this necessitates the use of exceptionally small time step sizes to achieve numerical convergence, if at all possible.
The use of small time steps may result in crucial problems when combined with "on the fly" electronic structure calculations of complex chemical systems.
There, even a single time step may become computationally expensive, in particular because of the calculations of forces and NAC vectors. 
This underscores the necessity of robust time integration approaches that allow for large time steps.

The present study introduces FSSH-2, an improved numerical time integration scheme for the FSSH method.
FSSH-2 incorporates a modified fewest switches hopping probability that utilizes time--overlap integrals for electronic amplitude propagation, a technique known as local diabatization \cite{Granucci2001}.
An essential feature of this approach is its numerical stability, as it operates independently of the potentially divergent NACs.
It is similar to the global flux hopping method \cite{Wang2014}, but stands apart in both its derivation and its efficient formulation.

We begin with a brief derivation of the conventional FSSH method in Sec. \ref{sec:FSSH}, highlighting its assumptions, numerical aspects, and established challenges.
In Sec. \ref{sec:FSSH2}, the FSSH-2 method is presented, along with a discussion of its theoretical advantages over FSSH. 
In Sec. \ref{sec:sim}, we present numerical experiments demonstrating the advantages of FSSH-2.

\section{FSSH Method} \label{sec:FSSH}

The traditional FSSH method propagates swarms of classical trajectories, that represent the nuclei, with associated electronic amplitudes that trigger hopping events to model non--adiabatic transitions.
We employ a fixed time step size denoted by $\Delta_t >0$, with time steps represented as $t_s = s \Delta_t$ for $s \in \N_0$.
We discuss the three fundamental tasks for advancing a given classical trajectory with position $q(t)\in\R^{d_{\nuc}}$ and momentum $p(t)\in\R^{d_{\nuc}}$ one time step further: classical propagation, evolution of the corresponding electronic amplitudes, and surface hopping.

\subsection{Classical Dynamics} \label{sec:cldy}
For each state, let the function $\theta_m (q,x)$ be the $m$-th adiabatic state, where $q \in \R^{d_{\nuc}}$ represents the nuclear coordinates and $x \in \R^{d_{\el}}$ the electronic coordinates.
Each adiabatic state is an eigenfunction corresponding to the $m$-th eigenvalue of the self--adjoint electronic Hamiltonian $H^{\el}(q)$ parametrically dependent on the coordinates $q$..
We assume that the adiabatic states are real--valued.
The eigenvalue equation for the electronic Hamiltonian can be written as
\begin{align*}
	H^{\el}(q) \theta_m (q,x) = E_m(q) \theta_m (q,x),
\end{align*}
where $E_m(q)$ defines the $m$-th adiabatic potential energy (hyper-)surface.
The classical dynamics of a trajectory occupying the adiabatic state $\theta_m$ is described by Hamilton's equations of motion:
\begin{align}
	\label{eq:cd}
	\begin{split}
		\dot q(t) &= \frac{p(t)}{M} \\
		\dot p(t) &= F_m(q(t))
		.
	\end{split}
\end{align}
Here, $M >0$ stands for the nuclear mass and $F_m(q) = -\nabla_q E_m(q) \in \R^{d_{\nuc}}$ represents the $m$-th adiabatic force
which can be computed by means of the Hellmann--Feynman theorem \cite{Hellmann1937,Feynman1939,Epstein1948},
\begin{align}
	\label{eq:HFforce}
	F_m(q) = 
	- \braket{\theta_m(q,\cdot) , \nabla_q H^{\el}(q)\theta_m(q,\cdot)}
	,
\end{align}
where $ \braket{\cdot,\cdot}$ represents the inner product of the electronic Hilbert space.

For propagating Eq. (\ref{eq:cd}), we used the Störmer--Verlet method \cite{Lasser2020} since it is an explicit, second--order, symplectic integrator.
Due to the singularity of the adiabatic force near eigenvalue crossings, we refrain from using a higher order method.

\subsection{Electronic Amplitudes} \label{sec:elecamp}
Let us denote by $L \in \N$ the total number of electronic states being considered. 
To compute hopping probabilities, FSSH propagates adiabatic amplitudes, denoted as $c^{\adi}(t) = \left( c^{\adi}_m(t) \right)_{m=1}^L \in \C^L$.
These amplitudes obey the following ordinary differential equation:
\begin{align}
	i \dot c^{\adi}(t) = H^{\adi}(t) c^{\adi}(t) 
	\label{eq:TDSE-electron-amplitudes}
	.
\end{align}
In this equation, the elements of the Hamiltonian matrix $H^{\adi}(t) = \left( H^{\adi}_{m,n}(t) \right)_{m,n=1}^L \in \C^{L \times L}$ are defined as
\begin{align*}
	H^{\adi}_{m,n}(t) = \delta_{m,n} E_m(q(t)) - \frac{i}{M} d^{m,n}(q(t))^T p(t) 
	,
\end{align*}
where $d^{m,n}(q) = \braket{\theta_m(q,\cdot) , \nabla_q\theta_n(q,\cdot)}$ represents the non--adiabatic coupling (NAC) vector.
Initially normalized amplitudes stay normalized for all times,$\|c^{\adi}(t)\| = 1$.
Similarly to the adiabatic force in Eq. (\ref{eq:HFforce}), the (extended) Hellmann--Feynman theorem provides a straightforward formula for the computation of the NAC vector for $E^m(q)\neq E^n(q)$:
\begin{align}
	d^{m,n}(q) = - \frac{\braket{\theta_m(q,\cdot) , \nabla_q H^{\el}(q)\theta_n(q,\cdot)}}{E_m(q)-E_n(q)}
	.
	\label{eq:nacHF}
\end{align}
However, propagating the adiabatic electronic amplitudes is significantly hampered by the divergence of the NAC vector, 
and the literature \cite{Hack1999, Jain2016b} recommends to introduce a second, smaller time step size for this task.

The direct propagation of Eq. (\ref{eq:TDSE-electron-amplitudes}) is typically carried out using the first term of the Magnus expansion \cite{Magnus1954a} \cite[Chapter IV.7]{Hairer2006} in conjunction with the trapezoidal rule:
\begin{align}
	c^{\adi}(t_{s+1}) 
	\approx 
	\exp \left( {- i \int \limits_{t_s}^{t_{s+1}} H^{\adi}(t) dt } \right) c^{\adi}(t_s)
	\approx 
	\exp \left( {- i \Delta_t  \frac{H^{\adi}(t_{s}) + H^{\adi}(t_{s+1})}{2} } \right) c^{\adi}(t_s)
	\label{eq:prop-ampl-TDSE-adi}
	.
\end{align}
The utilization of the trapezoidal rule here is advantageous as it circumvents the need to calculate adiabatic states at locations not associated with the defined time steps. 
As for the classical dynamics, a higher order approximation could lead to faster divergences for larger time step sizes.

\subsection{Surface Hopping} \label{sec:sh}
After propagating the electronic amplitudes, a trajectory may undergo a transition from its current state $m$ to a new state $n$. 
The FSSH hopping probability $P_{m,n}(t_s,t_{s+1})$ is derived by considering the electronic densities $\rho(t) = c^{\adi}(t) \overline{c^{\adi}(t)}^T \in \mathbb{C}^{L \times L}$ based on the electronic amplitudes. 
By means of Eq. (\ref{eq:TDSE-electron-amplitudes}), the rate of change of these densities are characterized by
\begin{align}
	\dot \rho_{m,m}(t) &= \frac{2}{M} \sum\limits_{\substack{n=1 \\ n \neq m}}^L \re{\rho_{m,n}(t)} \ d^{m,n}(q(t))^T p(t)
	\label{eq:LE-of-adi-pop}
	.
\end{align}
The FSSH hopping probability is obtained by tracking the outflux from state $m$ to all other states at time $t$ during a short time interval $\Delta_t$. 
Assuming no influx into state $m$ during this brief period, the rate of outflux is described as $\rho_{m,m}(t_{s}) - \rho_{m,m}(t_{s+1})$. 
The probabilities of staying at and leaving state $m$ are then given by:
\begin{align}
	P_{m,\stay}(t_{s},t_{s+1}) 
	= 
	\frac{\rho_{m,m}(t_{s+1})}{\rho_{m,m}(t_{s})}
	\quad \text{ and } \quad
	P_{m, \out}(t_{s},t_{s+1}) 
	= 
	\frac{\rho_{m,m}(t_{s}) - \rho_{m,m}(t_{s+1})}{\rho_{m,m}(t_{s})}
	\label{eq:outflux-hop-probability}
	.
\end{align}
This is known as the \textit{fewest switches} assumption after which the FSSH method is named. 
To compute the probability of hopping from state $m$ to another state $n \neq m$, we can express $P_{m,n}$ by inserting Eq. (\ref{eq:LE-of-adi-pop}) into (\ref{eq:outflux-hop-probability}):
\begin{align*}
	P_{m, \out}(t_{s},t_{s+1}) 
	= - \frac{\int_{t_{s}}^{t_{s+1}} \dot \rho_{m,m}(t) dt}{\rho_{m,m}(t_{s})}
	= - \frac{2}{M \rho_{m,m}(t_{s})} \sum\limits_{\substack{n=1 \\ n \neq m}}^L \int_{t_{s}}^{t_{s+1}} \re{\rho_{m,n}(t)} \ d^{m,n}(q(t))^T p(t) dt
	.
\end{align*}
The last equation formulates a hopping probability from state $m$ to $n$ as:
\begin{align}
	P_{m,n}(t_{s},t_{s+1}) &= - \frac{2}{M \rho_{m,m}(t_{s})} \int_{t_{s}}^{t_{s+1}} \re{\rho_{m,n}(t)} \ d^{m,n}(q(t))^T p(t) dt
	.
	\label{eq:FSSHhpcon}
\end{align}
By applying an Euler approximation to the integral, the original FSSH formula \cite{Tully1990} is reconstructed
\begin{align}
	P_{m,n}(t_s,t_{s+1}) &= - \frac{2 \Delta_t  \re{\rho_{m,n}(t_{s+1})}}{M \rho_{m,m}(t_s)} d^{m,n}(q(t_{s+1}))^T p(t_{s+1})
	\label{eq:FSSHhp}
	,
\end{align}
where negative values are treated as zeros. 
Note that $P_{m,n}$ depends on the NAC, thus inheriting its possible divergence issue. 

\bigskip 

For a hopping event from state $m$  to another state $n$, the process involves a uniformly distributed random number $r \in [0,1]$. 
The state \( n \) is determined by the condition:
\begin{align*}
	\sum_{\substack{k=1\\ k \neq m}}^{n-1} P_{m,k}(t_{s},t_{s+1}) \leq r < \sum_{\substack{k=1\\ k \neq m}}^{n} P_{m,k}(t_{s},t_{s+1}) .
\end{align*}
If no such $n$ satisfies this criterion, no hopping is performed and the trajectory remains in state $m$.
This random hopping criterion together with initial random sampling constitute the stochastic nature of the FSSH method.

\section{FSSH-2 Method} \label{sec:FSSH2}
The above numerical issues have led us to explore alternative and more efficient approaches to refine the FSSH method. 
The novel FSSH-2 presents two modifications to the original FSSH method: the propagation of adiabatic amplitudes through local diabatization and the redefinition of FSSH hopping probabilities.

\subsection{Local Diabatization Method for Propagating Adiabatic Amplitudes}
In the spirit of the local diabatization method \cite{Granucci2001}, we use a diabatic representation, which often offers smoother conditions.
For the while being, we assume that a diabatic potential energy matrix $V^{\dia}(q) \in \R^{L \times L}$ is given but later on it will turn out that an adiabatic--to--diabatic transformation is not required.
Instead of considering the adiabatic amplitudes, we work in a diabatic representation with diabatic amplitudes that evolve according to
\begin{align}
	i \dot c^{\dia}(t) = H^{\dia}(t) c^{\dia}(t) 
	\label{eq:TDSE-electron-amplitudes-dia}
	,
\end{align}
where the diabatic Hamiltonian consists solely of the diabatic potential energy matrix, $H^{\dia}(t) = V^{\dia}(q(t))$.
The diabatic amplitudes are related to the adiabatic amplitudes through the diagonalization of $V^{\dia}(q)$
\begin{align}
	V^{\dia}(q) = \overline{U(q)}^T E(q) U(q),
	\label{eq:diagdia}
\end{align}
where $E(q)=\diag \left(E_1(q),...,E_L(q)\right) \in \R^{L \times L}$ denotes the adiabatic potential energy matrix, which is diagonal.
Consequently, the unitary matrix $U(q(t)) \in \R^{L \times L}$ serves as the transformation matrix from diabatic to adiabatic representation. 
This implies that the electronic amplitudes in the diabatic and adiabatic representations are related as follows:
\begin{align}
	c^{\adi}(t) = U(q(t)) c^{\dia}(t)
	\label{eq:diaadirel}
	.
\end{align}
Note that we will show below that the transformation matrices $U$ will not be used in the FSSH-2 scheme.
Analogously to Eq. (\ref{eq:prop-ampl-TDSE-adi}), we can propagate Eq. (\ref{eq:TDSE-electron-amplitudes-dia}) using the first term of the Magnus expansion but this time  in conjunction with the Euler method
\begin{align*}
	c^{\dia}(t_{s+1}) \approx \exp \left( {- i \int \limits_{t_s}^{t_{s+1}} V^{\dia}(q(t)) dt } \right) c^{\dia}(t_s)
	\approx 
	\exp \left( {- i \ V^{\dia}(q(t_{s+1})) \ \Delta_t } \right) c^{\dia}(t_s) 
	.
\end{align*}
The Euler method is beneficial here since the diagonalization of $V^{\dia}(q(t))$ is inherently involved in the force calculation for classical dynamics when using a diabatic model.
Together with Eq. (\ref{eq:diagdia}), this leads to:
\begin{align*}
	c^{\dia}(t_{s+1}) 
	&\approx 
	\exp \left( {- i \ V^{\dia}(q(t_{s+1})) \ \Delta_t } \right) c^{\dia}(t_s) 
	\\
	&= 
	\overline{U(q(t_{s+1}))}^T \exp \left( {- i \ E(q(t_{s+1})) \ \Delta_t } \right) U(q(t_{s+1})) \ c^{\dia}(t_s)
	.
\end{align*}
Transforming the diabatic amplitudes back into the adiabatic representation by using Eq. (\ref{eq:diaadirel}), we obtain the following propagator for the adiabatic amplitudes:
\begin{align}
	c^{\adi}(t_{s+1})
	&\approx \exp \left( {- i \ E(q(t_{s+1})) \ \Delta_t } \right) \Gamma(t_s,t_{s+1}) \ c^{\adi}(t_s)
	\label{eq:loc-dia-prop}
\end{align}
with $\Gamma(t_s,t_{s+1}) = U(q(t_{s+1})) \overline{U(q(t_{s}))}^T \in \R^{L\times L}$. 
Evaluating the remaining matrix exponential is straightforward, as $E(q)$ is a diagonal matrix. 
Interestingly, this propagator remains valid even in the absence of a known diabatic framework. 
The matrix elements of $\Gamma(t_s,t_{s+1})$ are defined as 
\begin{align*}
	\Gamma_{m,n}(t_s,t_{s+1}) = \braket{\theta_m(q(t_{s+1}),\cdot) ,\theta_n(q(t_s),\cdot)}
	,
\end{align*}
where the $\theta$ are the adiabatic states defined in Sec. \ref{sec:cldy}.
In Ref. \cite{Fabiano2008}, these matrix elements are referred to as time--overlaps between adiabatic states.
From the point of view of numerical calculations, they have the advantage that their absolute value is bounded by one. 
Thus, Eq. (\ref{eq:loc-dia-prop}) results in a propagator of the adiabatic states that only depends on the adiabatic potential energies, $E$, and time--overlaps, $\Gamma$, of adiabatic states.
Hence, this propagator offers several advantages in addressing the numerical challenges discussed for Eq. (\ref{eq:prop-ampl-TDSE-adi}).
First, it eliminates the need to consider the NAC vectors, thus ensuring there are no divergent quantities in the calculations and computational costs do not scale with $\bigo(L^4)$, as demonstrated in our previous work \cite{Schmidt2019}. 
In fact, the computational costs of $\Gamma$ are comparable to those of matrix multiplication, scaling with $\bigo(L^2)$. 
Second, the implementation of FSSH-2 is straightforward and all operations in Eq. (\ref{eq:loc-dia-prop}) can be trivially parallelized across all trajectories.
Thus, the local diabatization propagator offers a simple, cost--effective, and parallelizable method for propagating adiabatic amplitudes.

\subsection{FSSH-2 Hopping Formula} \label{sec:fssh2-HF}
Another challenge is to derive a hopping formula that can harness the advantages of the local diabatization propagator fully. 
The traditional FSSH hopping formula exhibits the stability issues mentioned in Sec. \ref{sec:elecamp} and \ref{sec:sh}.
To tackle this, we revisit the outflow probability in Eq. (\ref{eq:outflux-hop-probability}). 
By first considering the case of only two electronic states ($L=2$), the outflow probability characterizes the hopping probability from state $m$ to the other state $n \neq m$.
Utilizing the normalization of our electronic densities, $\rho_{m,m}(t) + \rho_{n,n}(t) = 1$, we can express the outflow probability in terms of state $n$:
\begin{align}
	P_{m, \out}(t_s,t_{s+1})
	=
	\frac{\rho_{m,m}(t_s) - \rho_{m,m}(t_{s+1})}{\rho_{m,m}(t_s)}
	=
	\frac{\rho_{n,n}(t_{s+1}) - \rho_{n,n}(t_s)}{\rho_{m,m}(t_s)}
	.
	\label{eq:fssh22state}
\end{align}
This hopping formula is straightforward to evaluate, given that all its elements are densities, bounded by 1. 
Notably, this formula is not only more concise than the FSSH hopping formula in Eq. (\ref{eq:FSSHhp}) but it is also independent of NACs. 
Consequently, it provides a simple approach to circumvent potential challenges of NACs.

We extend this approach to an arbitrary number of electronic states $L \in \mathbb{N}$. 
The normalization condition for electron densities can be expressed as
\begin{align*}
	\sum_{n=1}^L \rho_{n,n} (t) = 1
	\quad
	\Leftrightarrow
	\quad
	\rho_{m,m} (t) = 1 - \sum_{ \substack{n=1 \\ n \neq m}}^L \rho_{n,n}(t).
\end{align*}
Substituting this into Eq. (\ref{eq:outflux-hop-probability}), we obtain the expression
\begin{align*}
	P_{m, \out}(t_s,t_{s+1}) = \sum_{ \substack{n=1 \\ n \neq m}}^L \frac{\rho_{n,n} (t_{s+1}) - \rho_{n,n} (t_s)}{\rho_{m,m}(t_s)}.
\end{align*}
The individual terms in this sum, $\frac{\rho_{n,n} (t_{s+1}) - \rho_{n,n} (t_s)}{\rho_{m,m}(t_s)}$, indicate a hopping probability between state $m$ and $n$, analogous to Eq. (\ref{eq:fssh22state}) for the case of two electronic states.
However, for $L > 2$, the physical interpretation of these terms might be incorrect in certain scenarios.
For instance, when $\rho_{n,n} (t_{s+1}) - \rho_{n,n} (t_s)>0$ and the outflow of state $m$ is zero, it could incorrectly suggest a hopping from $m$ to $n$.
Therefore, we introduce $P_{m, \out}$ as an upper bound and define the FSSH-2 hopping probability as
\begin{align}
	\tilde P_{m,n}(t_s,t_{s+1}) = \min \left\lbrace P_{m, \out}(t_s,t_{s+1}) , \frac{\rho_{n,n} (t_{s+1}) - \rho_{n,n} (t_s)}{\rho_{m,m}(t_s)} \right\rbrace 
	\label{eq:FSSH2-hopping}.
\end{align}
As mentioned for the case $L=2$, it addresses some of the issues encountered with the traditional FSSH hopping formula. 
Based solely on the densities derived from adiabatic amplitudes, $\tilde P_{m,n}$ circumvents the issues related to the divergent NAC vectors. 
Additionally, the computation of this refined hopping formula is simple, enhancing its practicality. 
With the absence of approximations in its formulation, the precision of $\tilde P_{m,n}$ is solely connected to the accuracy of the propagation of the adiabatic amplitudes.

\subsection{Limitations of the FSSH-2 hopping probability}
The FSSH-2 hopping probability still presents certain shortcomings. 
Like FSSH, the probabilities do not form a valid probability distribution. 
Another concern is that the upper bound in Eq. (\ref{eq:FSSH2-hopping}) may not be sufficient to prevent incorrect transfers between states that are not coupled but have transfers at the same time. 
Imagine a scenario with four states where the first and last two states cross each other simultaneously, but there is a substantial gap between states 2 and 3. 
If there is population on all states, it may occur that a hopping event from state 3 to 2 occurs.

\section{Simulations} \label{sec:sim}

\begin{table}[t]
	\centering
	\resizebox{\textwidth}{!}{%
		\begin{tabular}{>{\centering\arraybackslash} m{2cm} >{\centering\arraybackslash} m{6cm} >{\centering\arraybackslash} m{6cm} >{\centering\arraybackslash} m{6cm}} \toprule \toprule 
Model & Difference between quantum solution and FSSH & Difference between quantum solution and FSSH-2 & Difference between FSSH \newline and FSSH-2 \\  
 \cmidrule(lr){1-1} \cmidrule(lr){2-2} \cmidrule(lr){3-3} \cmidrule(lr){4-4} 
Tully 1 & 0.004 & 0.003 & 0.001 \\ 
Tully 2 & 0.008 & 0.004 & 0.007 \\ 
2D LVC & 0.015 & 0.012 & 0.014 \\ 
2D Well & 0.023 & 0.021 & 0.004 \\ 
Model X & 0.007 & 0.010 & 0.003 \\ 
\bottomrule \bottomrule 
\end{tabular}

	}
	\caption{
		Comparison of the quantum reference solution to converged results of FSSH and FSSH-2 for each model, evaluated at our highest chosen (256) substep number.
		The numbers showcase the mean absolute population differences of the second adiabatic state.
	}
	\label{tbl:comparison_convergence}
\end{table}

The FSSH-2 method offers a straightforward yet promising redefinition of the traditional FSSH method, showing potential for enhanced numerical efficiency, which needs further validation across various scenarios.
To assess the performance of the two methods, we explored a variety of diabatic models of non--adiabatic dynamics in molecular systems: 
the Tully single and double crossing models \cite{Tully1990}, the two--dimensional linear vibronic coupling (2D LVC) model \cite{Gherib2016}, the two--dimensional well \cite{Shenvi2011}, and Model X \cite{Subotnik2011}.
While the latter one incorporates three electronic states, the other models feature two electronic states.
This exhaustive approach is in line with the recommended test suite for non--adiabatic algorithms by Nelson et al. \cite{Nelson2020b}.

Our simulations were carried out using version 7.2.0 of the WavePacket MATLAB software package \cite{Schmidt2017, Schmidt2017a, Schmidt2019}, which is designed to treat fully quantum--mechanical (grid based) and quantum--classical (trajectory based) simulations on an equal footing. 
Given that the use of grid--based representations of wave functions is restricted to low dimensional systems, its quantum--classical counterpart offers an efficient alternative for simulations involving higher dimensional systems.
Within WavePacket, the main time steps can be divided into an arbitrary number of substeps, which are then used as our previously defined (propagation) time steps $\Delta_t$.
Therefore, by keeping the main time steps fixed for a model, we can vary $\Delta_t$ by varying the number of substeps.
This also enables us to easily maintain consistency across the models by selecting the same substep number.

In our simulations, we chose the simulation time such that the classical trajectories could traverse the (avoided) crossings up to two times. 
The surface hopping methods used swarms of 10,000 trajectories. 
Following each hopping event, the momentum was rescaled along the NAC vectors, the direction of which can be efficiently computed without singularity issues by evaluating only the numerator in Eq. (\ref{eq:nacHF}). 
The main time step sizes were selected based on the limits of energy conservation of the classical dynamics propagator. 
To compare our results to a full quantum solution, we also performed a wave function simulation using a second--order Strang-Marchuk operator splitting scheme with the FFT grid representation~\cite{Askar1978,Kosloff1983,Leforestier1991}. 
Further details for each simulation set--up can be found in the appendix.

\begin{figure}[t]
	\centering
	\includegraphics[width=0.329\linewidth]
	{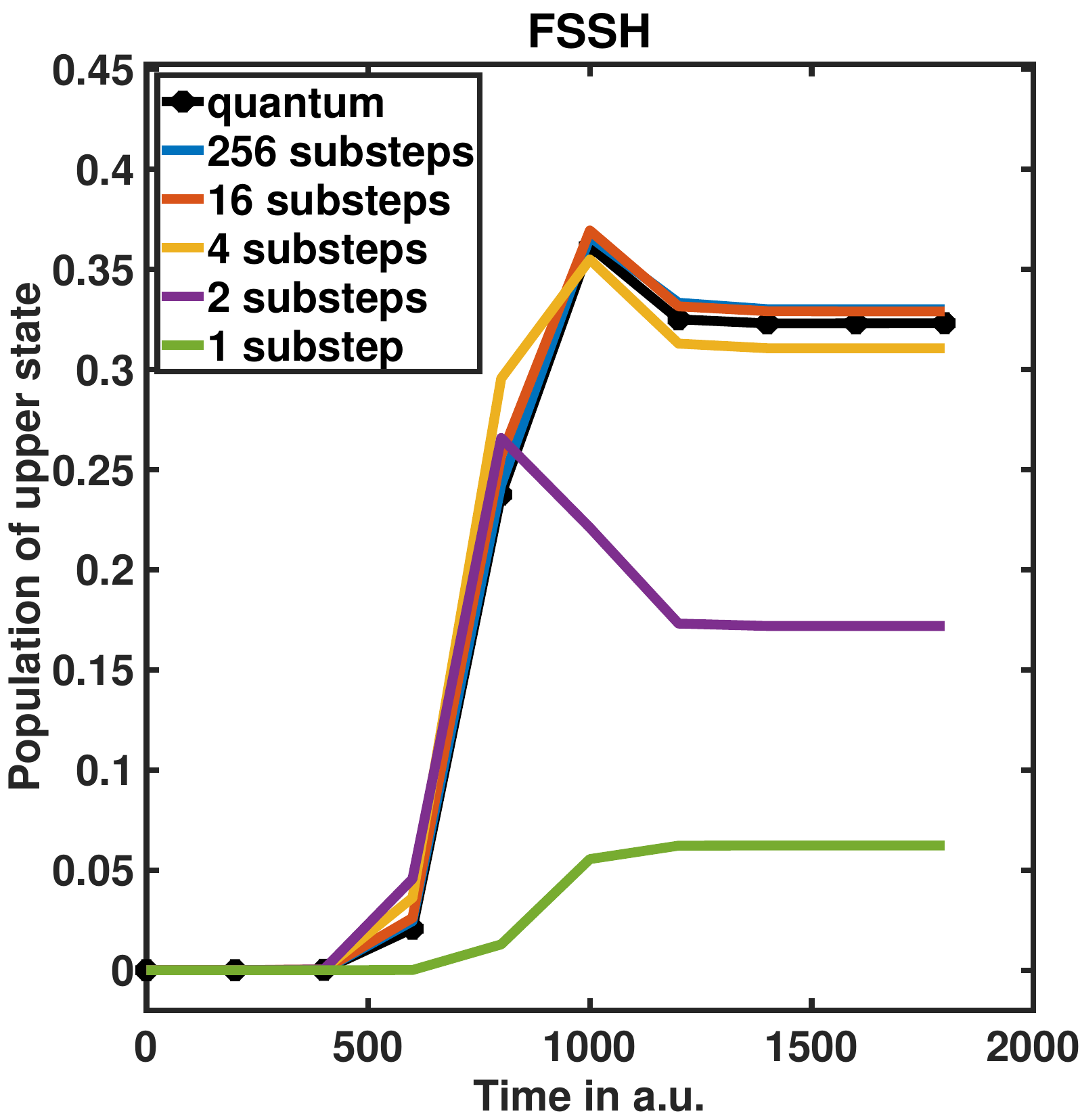}
	\includegraphics[width=0.329\linewidth]
	{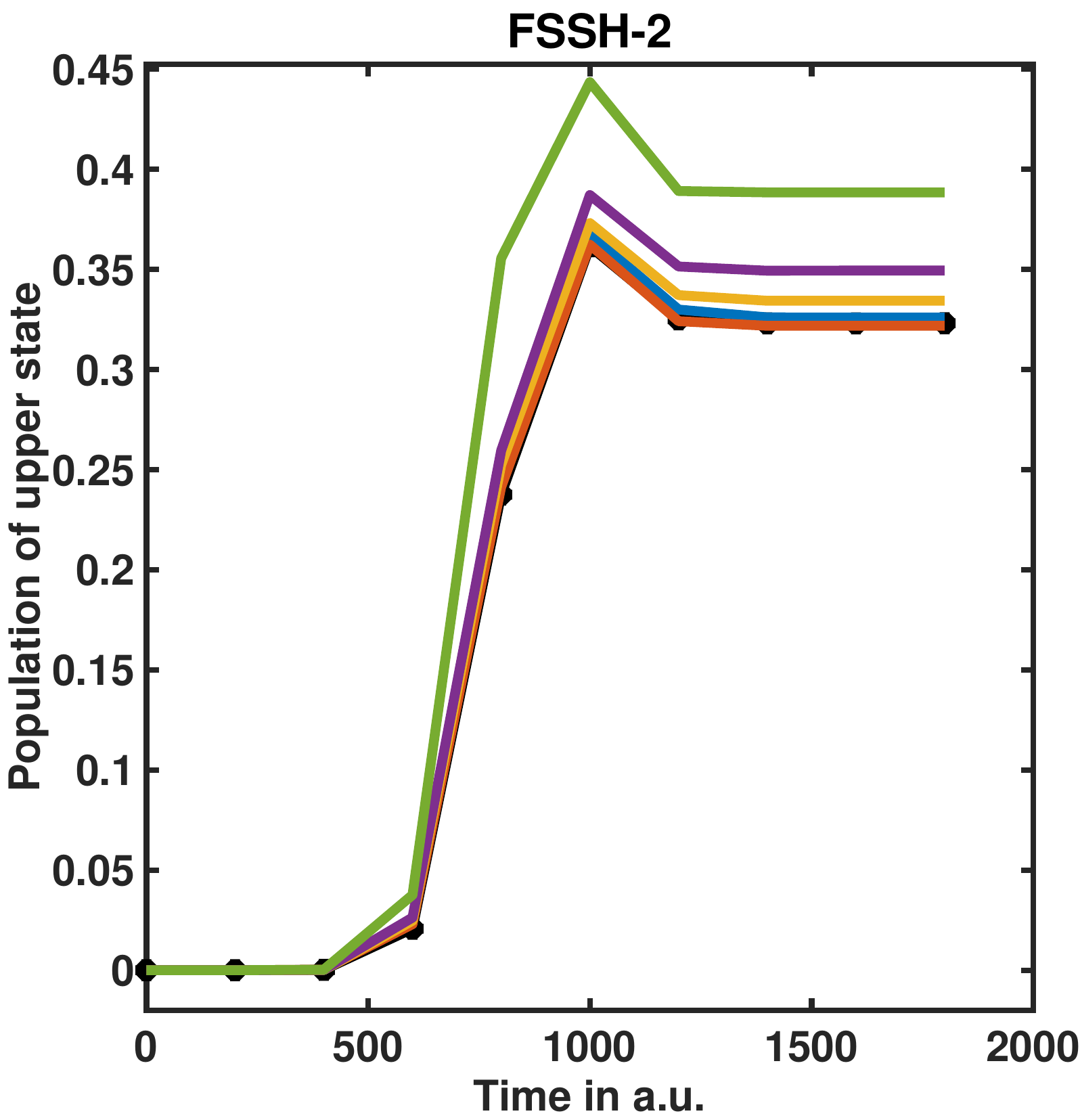}
	\includegraphics[width=0.329\linewidth]
	{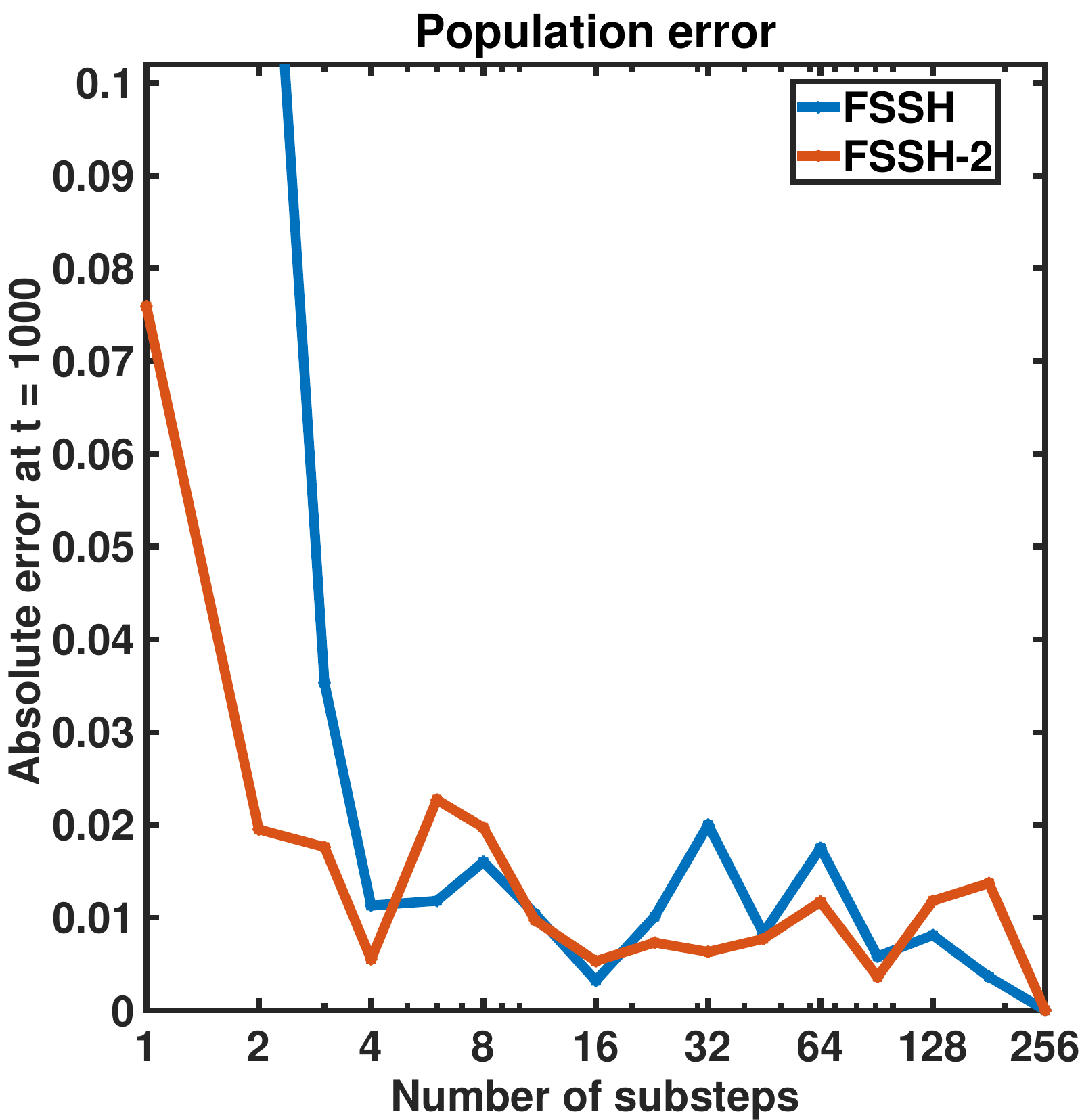}
	\caption{
		Population of the second adiabatic state for the Tully 1 model using FSSH (left) and FSSH-2 (center) for different numbers of substeps.
		The right panel shows the absolute population error after the initial hopping event for both methods.
	}
	\label{fig:plotmethodconvT1}
\end{figure}

\begin{figure}[t]
	\centering
	\includegraphics[width=0.329\linewidth]
	{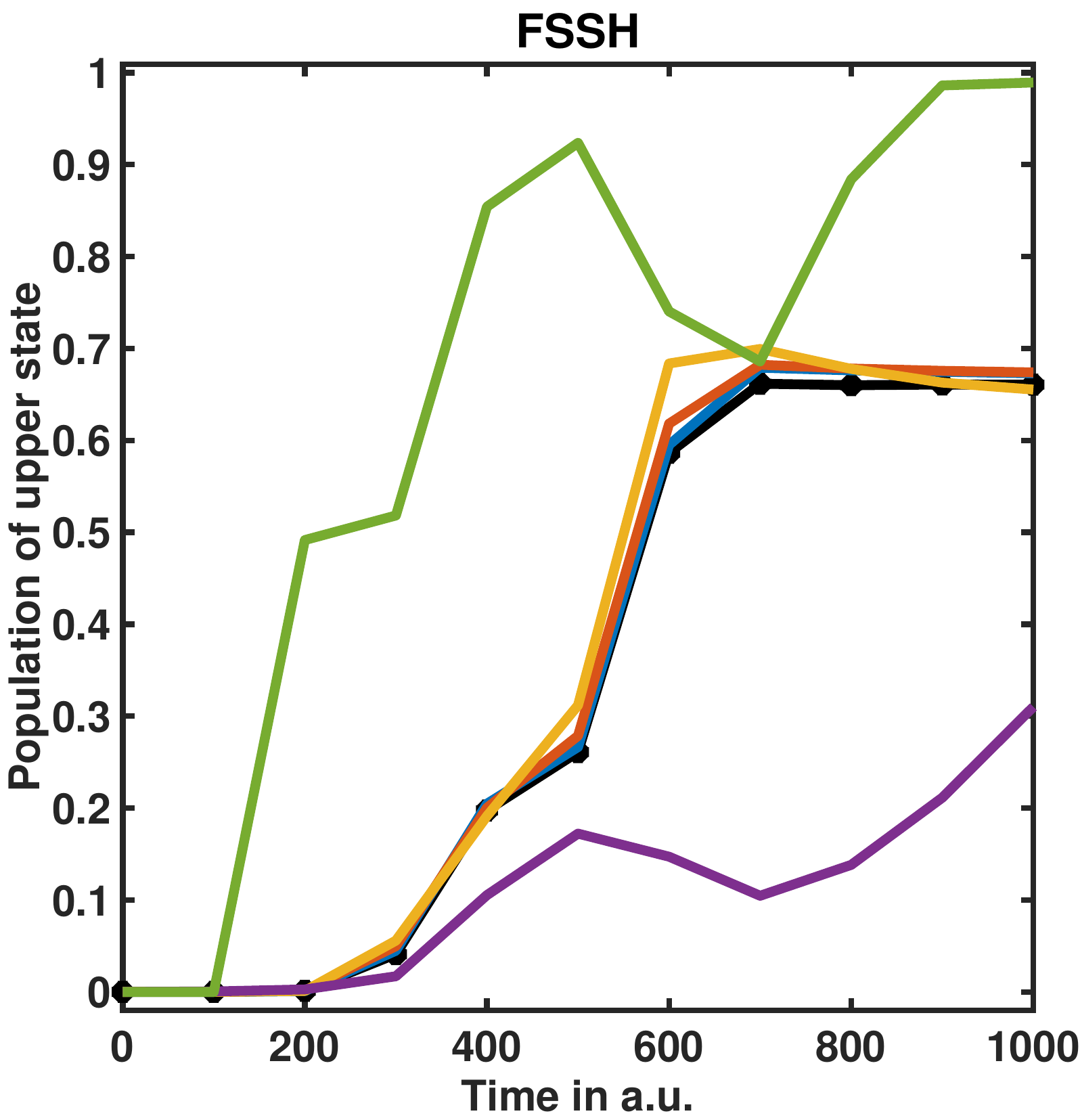}
	\includegraphics[width=0.329\linewidth]
	{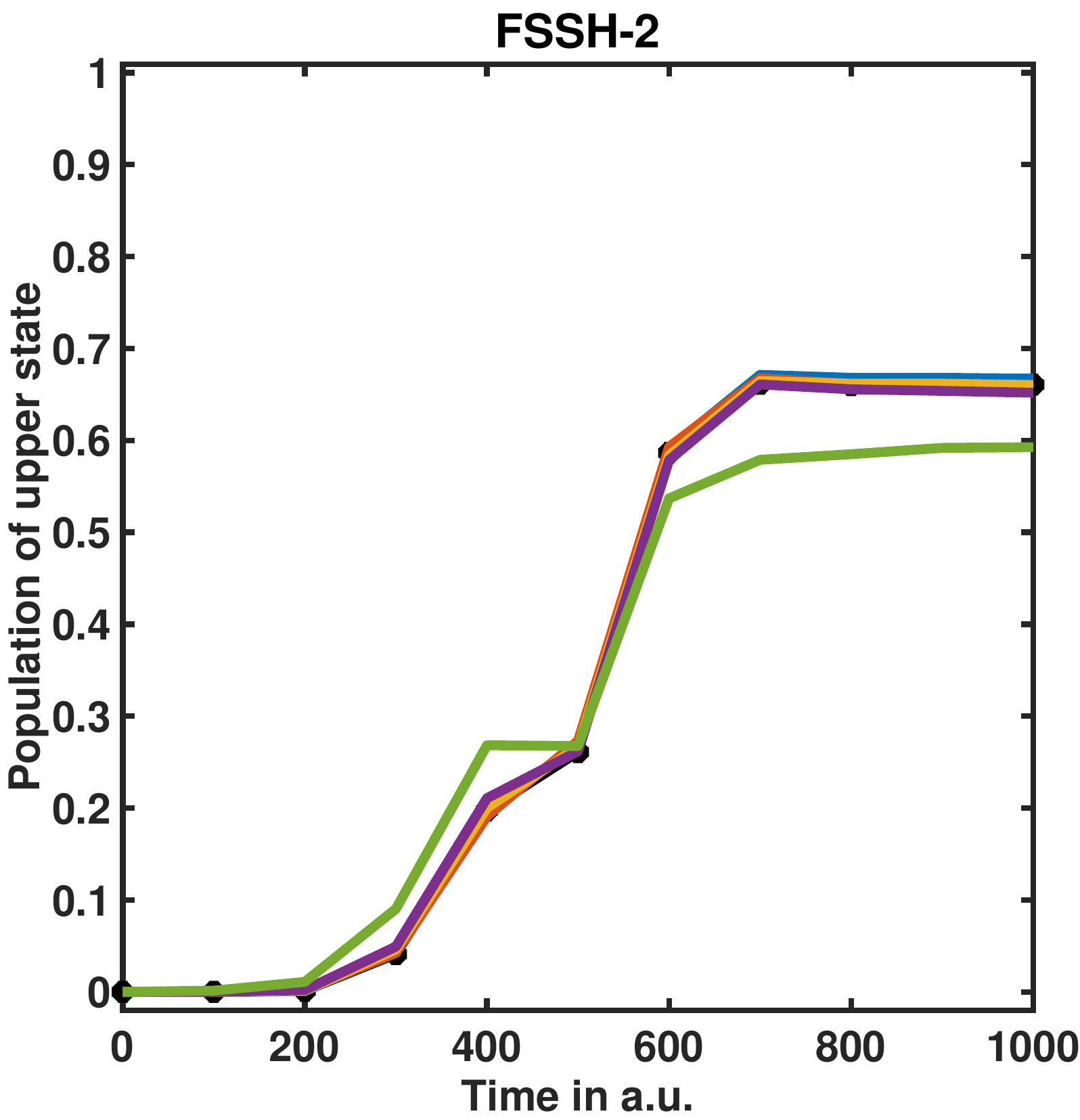}
	\includegraphics[width=0.329\linewidth]
	{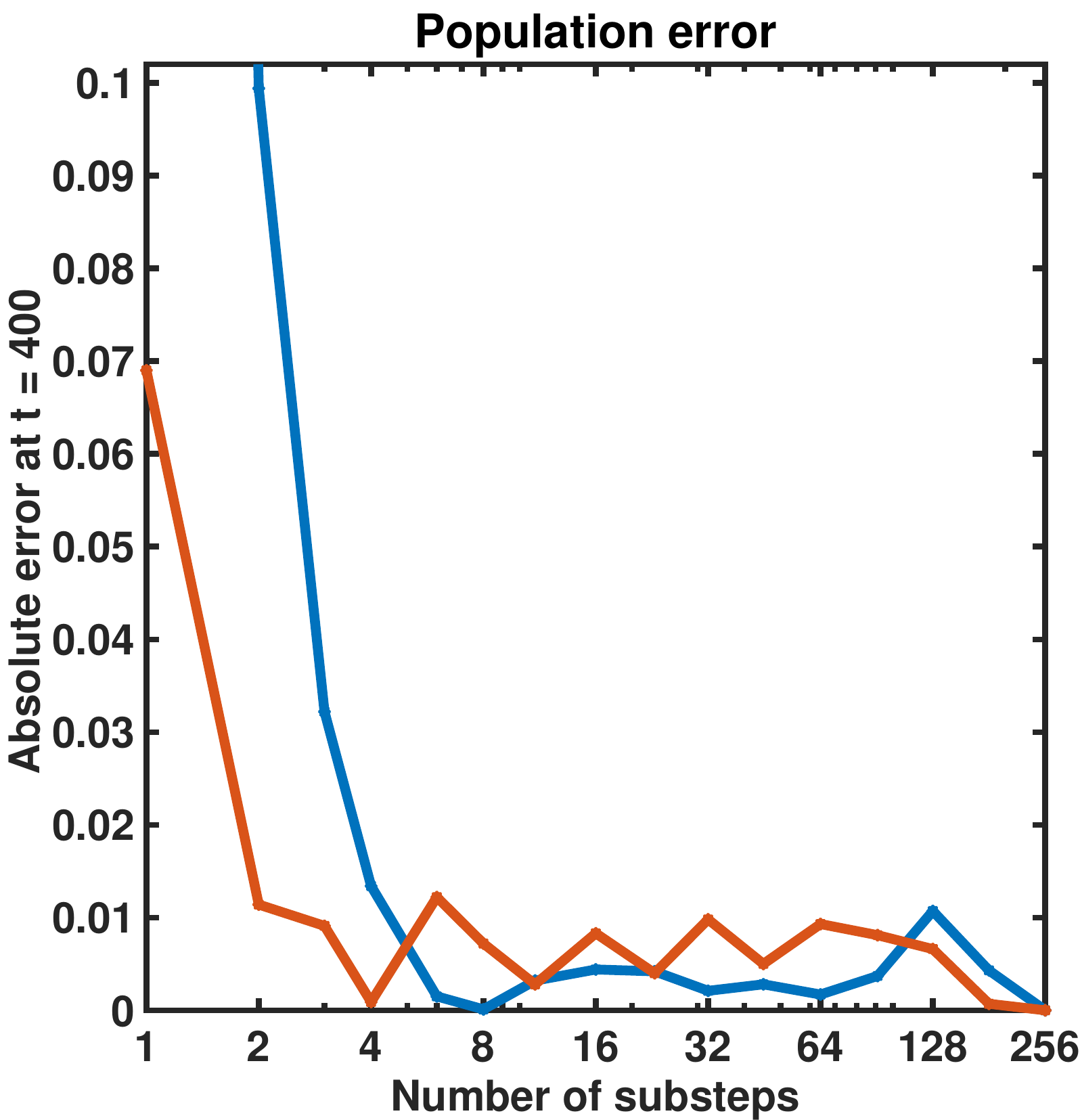}
	\caption{
		Same as Fig. \ref{fig:plotmethodconvT1} but for the Tully 2 model.
	}
	\label{fig:plotmethodconvT2}
\end{figure}

\begin{figure}[t]
	\centering
	\includegraphics[width=0.329\linewidth] {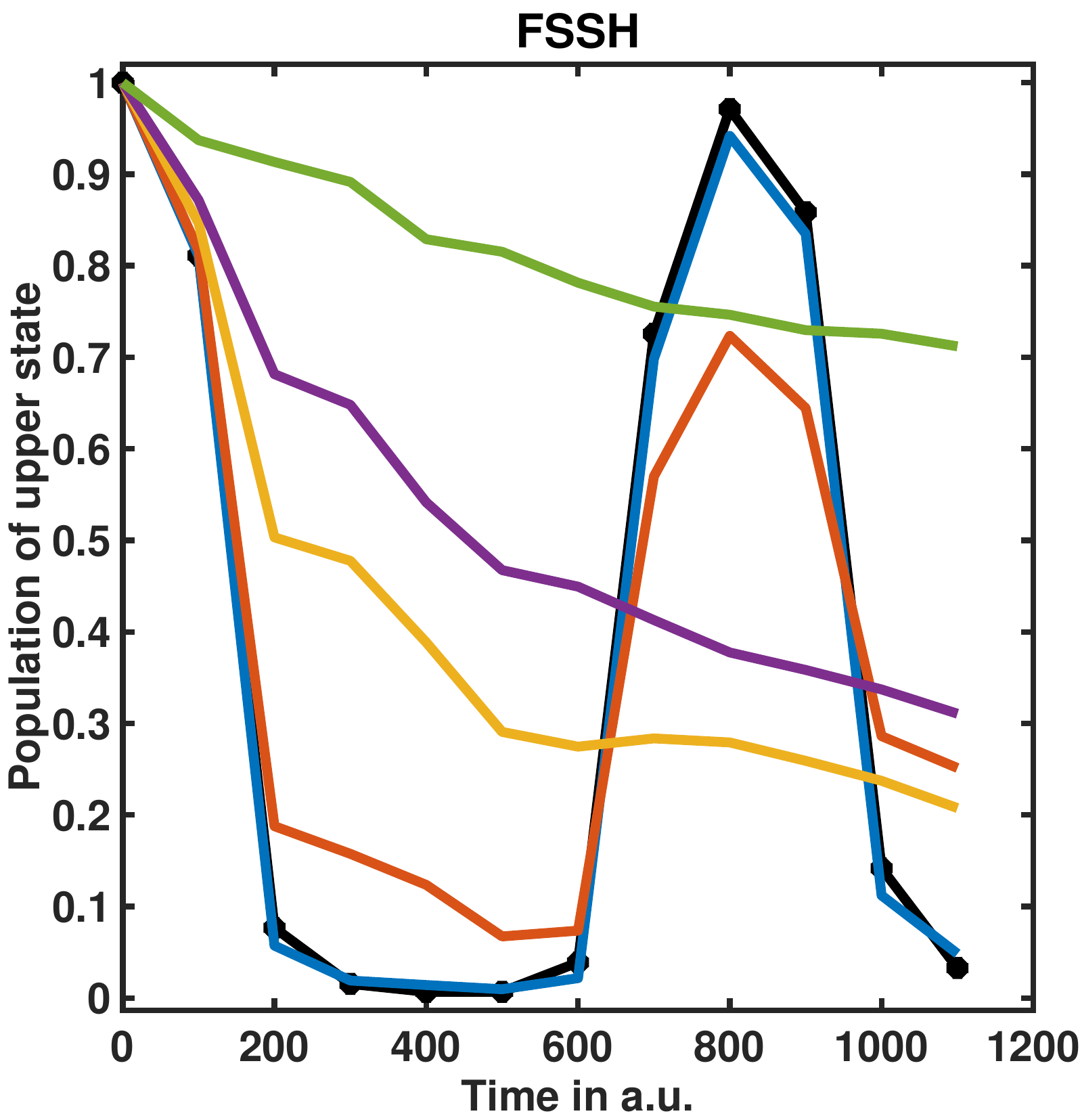}
	\includegraphics[width=0.329\linewidth] {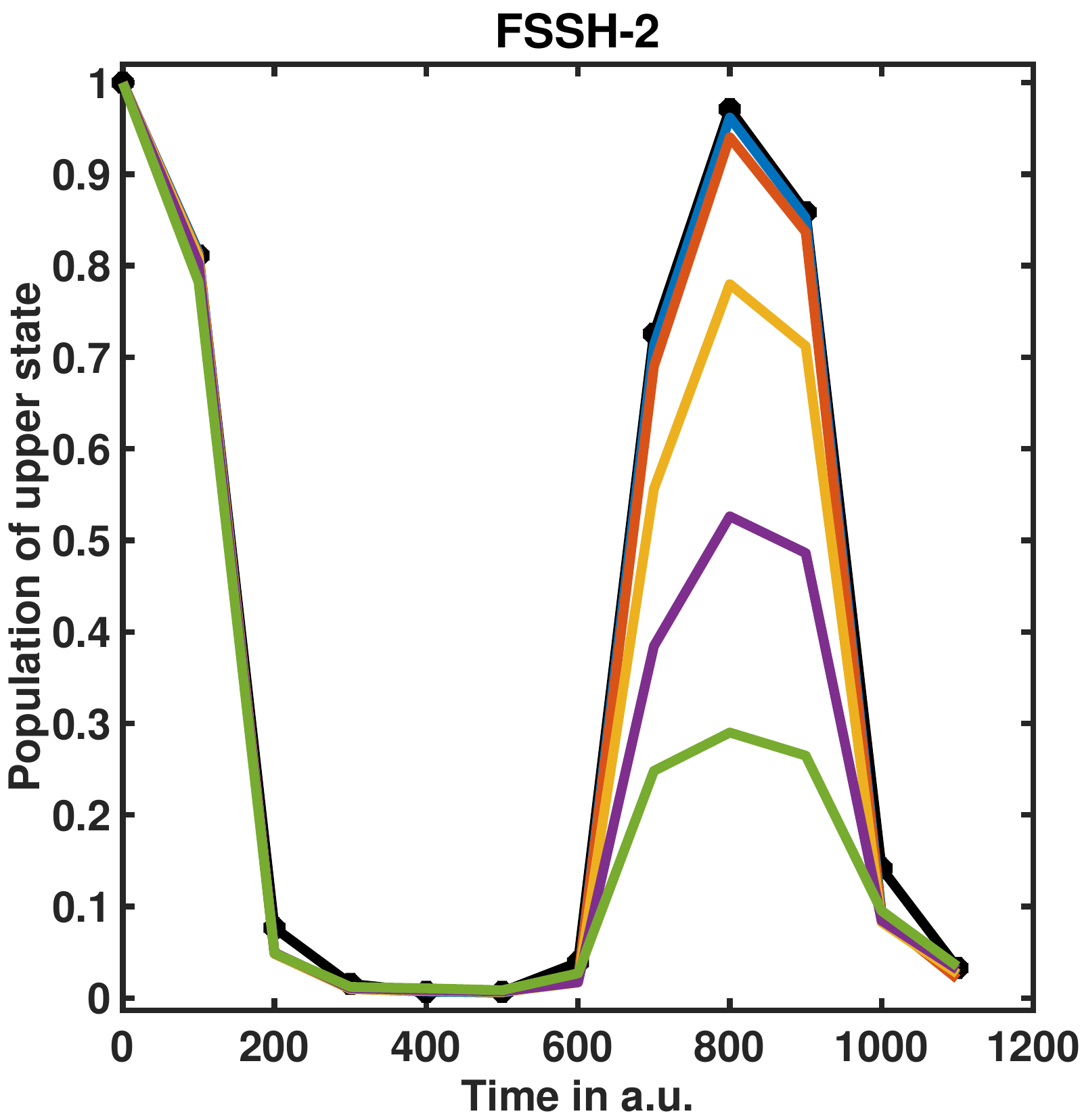}
	\includegraphics[width=0.329\linewidth]
	{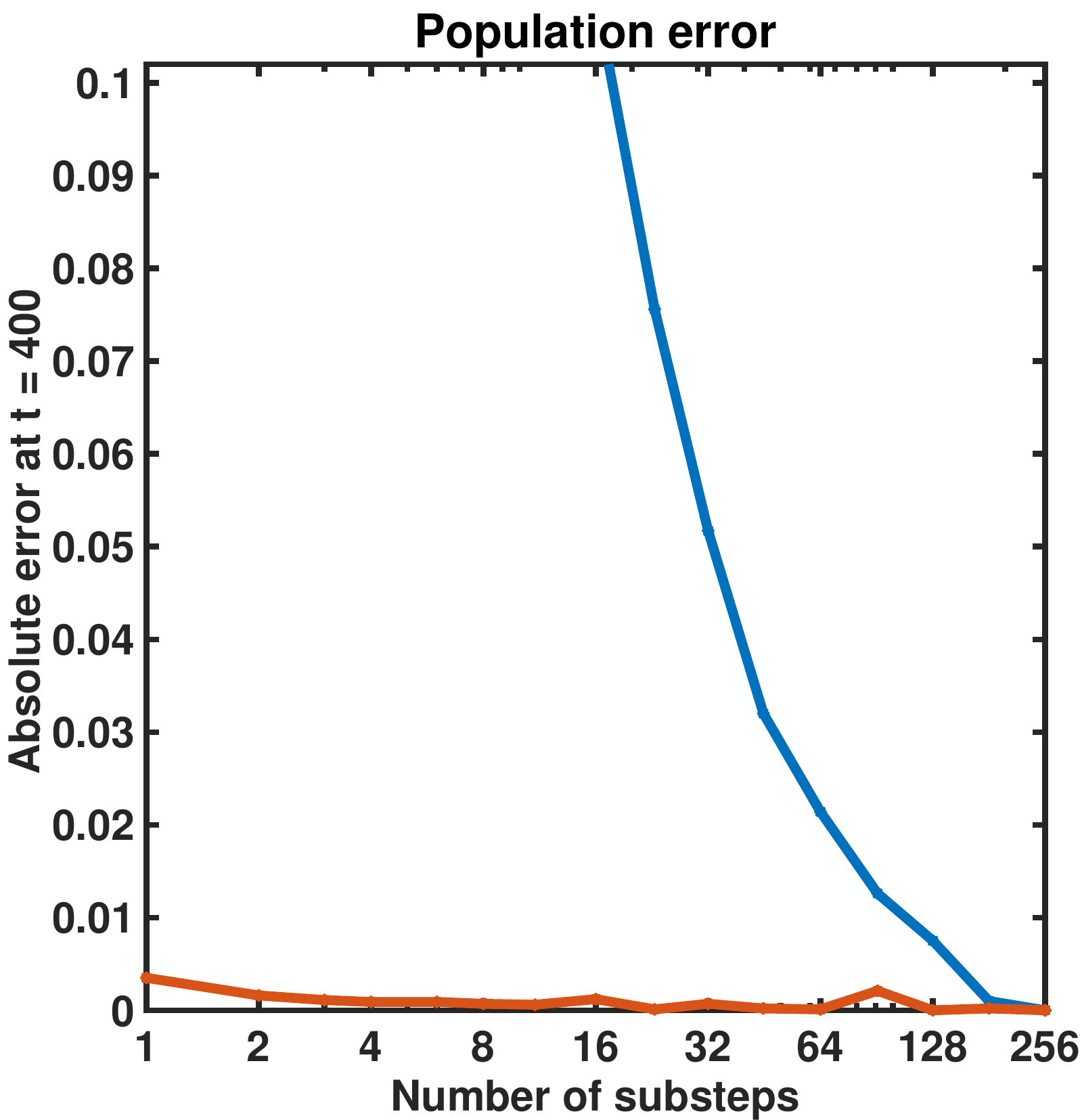}
	\caption{
		Same as Fig. \ref{fig:plotmethodconvT1} but for the 2D LVC model.
	}
	\label{fig:plotmethodconvBMA}
\end{figure}

\begin{figure}[t]
	\centering
	\includegraphics[width=0.329\linewidth] {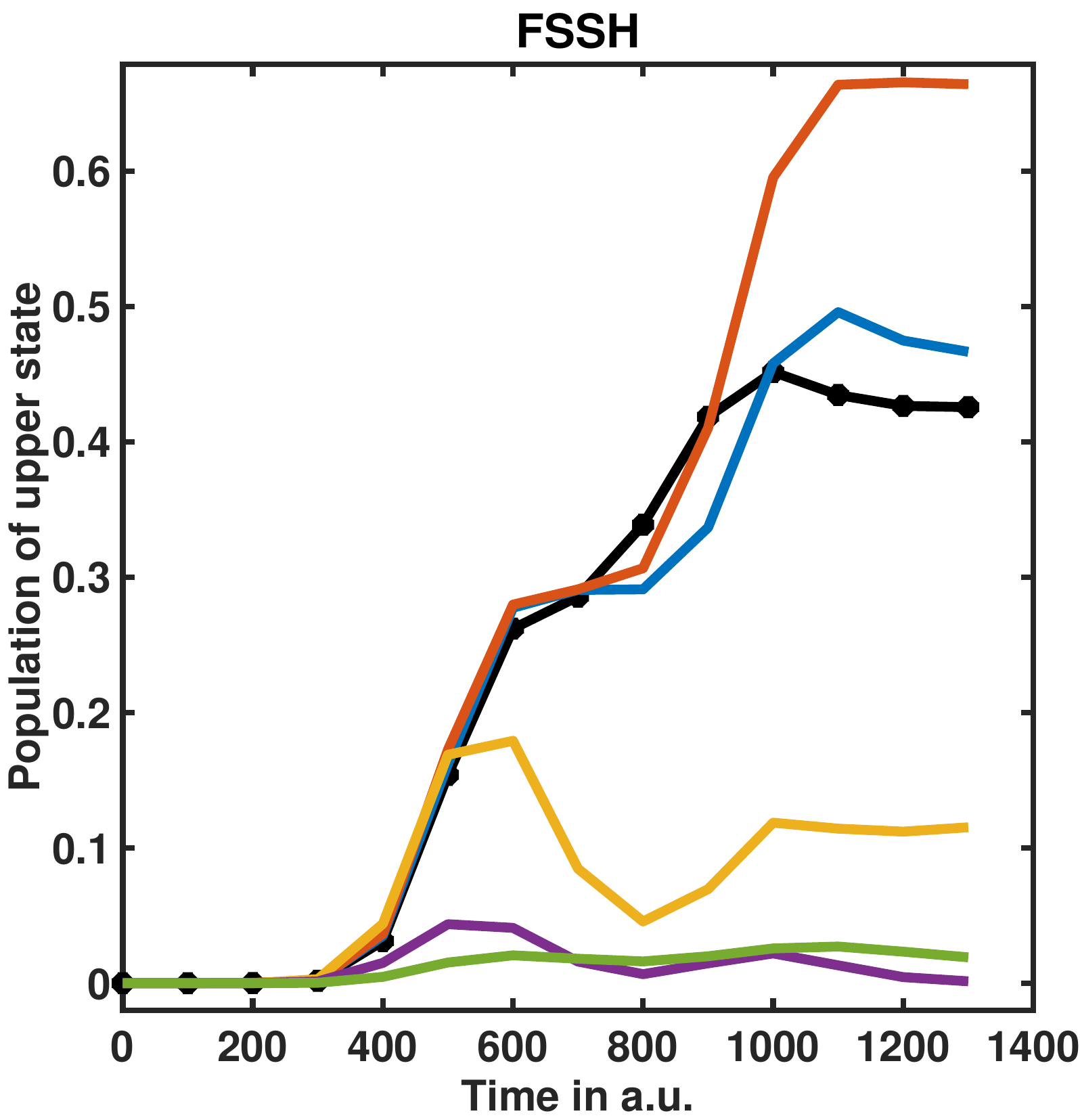}
	\includegraphics[width=0.329\linewidth] {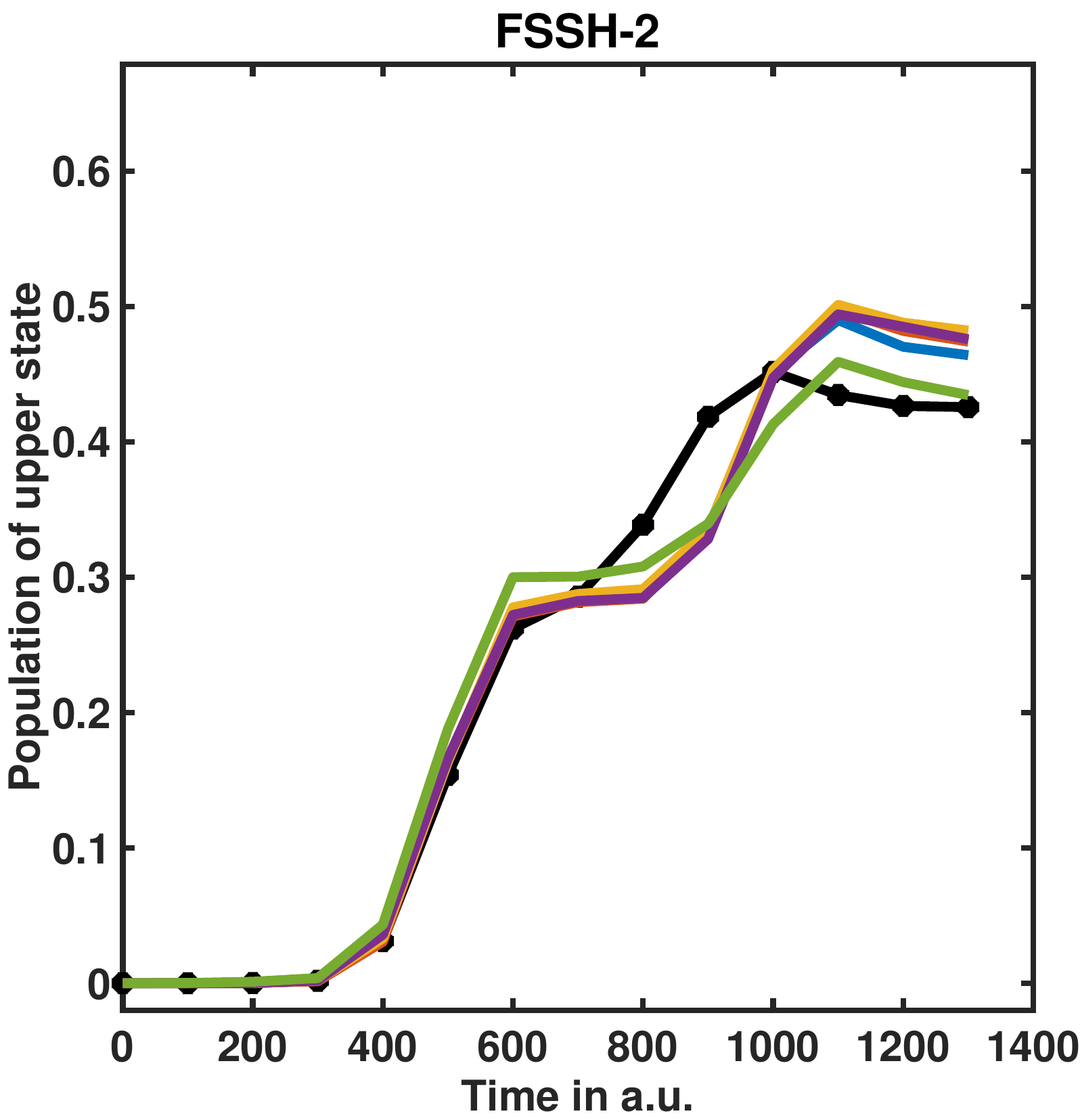}
	\includegraphics[width=0.329\linewidth] 
	{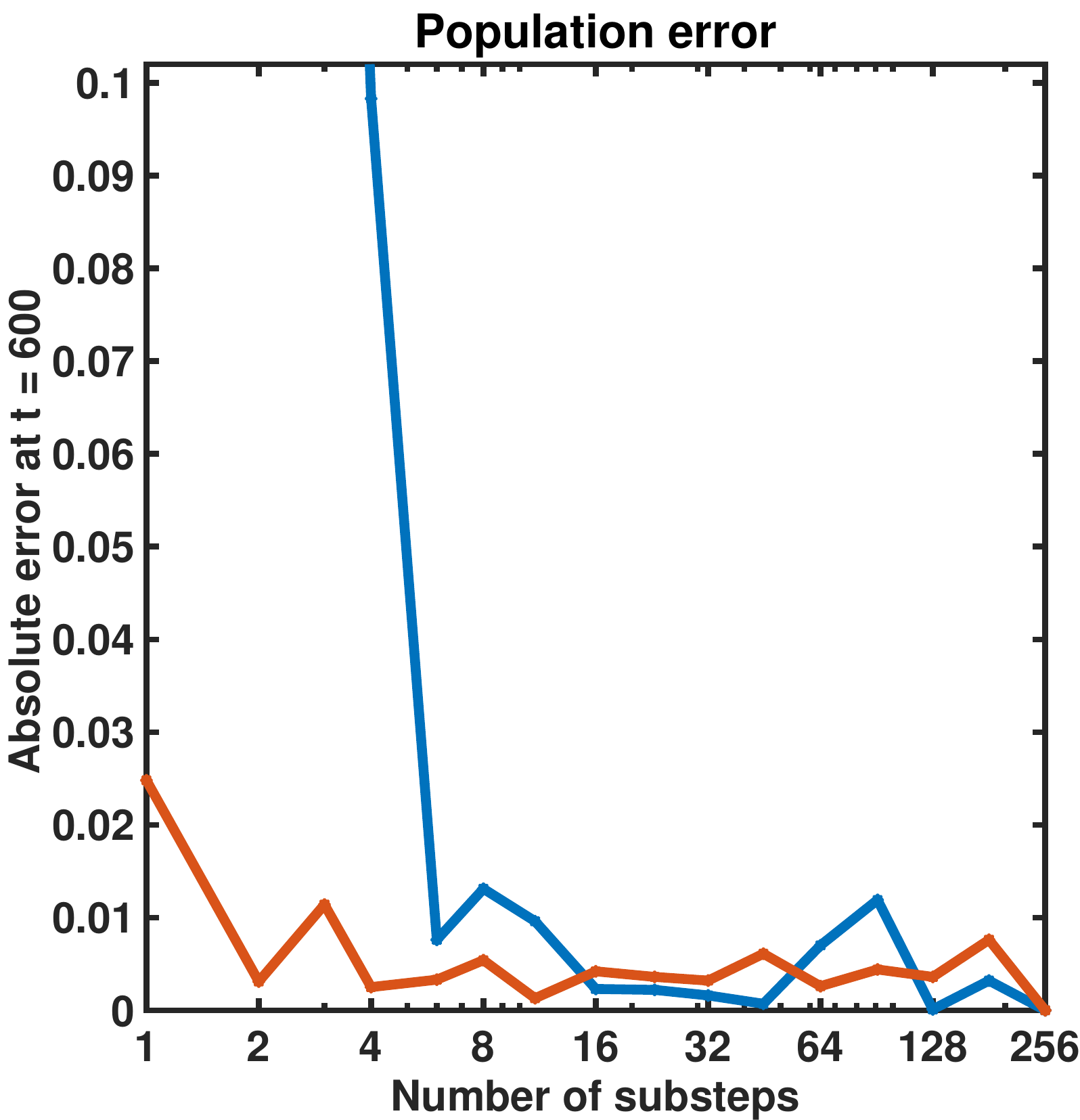}
	\caption{
		Same as Fig. \ref{fig:plotmethodconvT1} but for the 2D Well model.
	}
	\label{fig:plotmethodconvWell}
\end{figure}

\begin{figure}[t]
	\centering
	\includegraphics[width=0.329\linewidth] {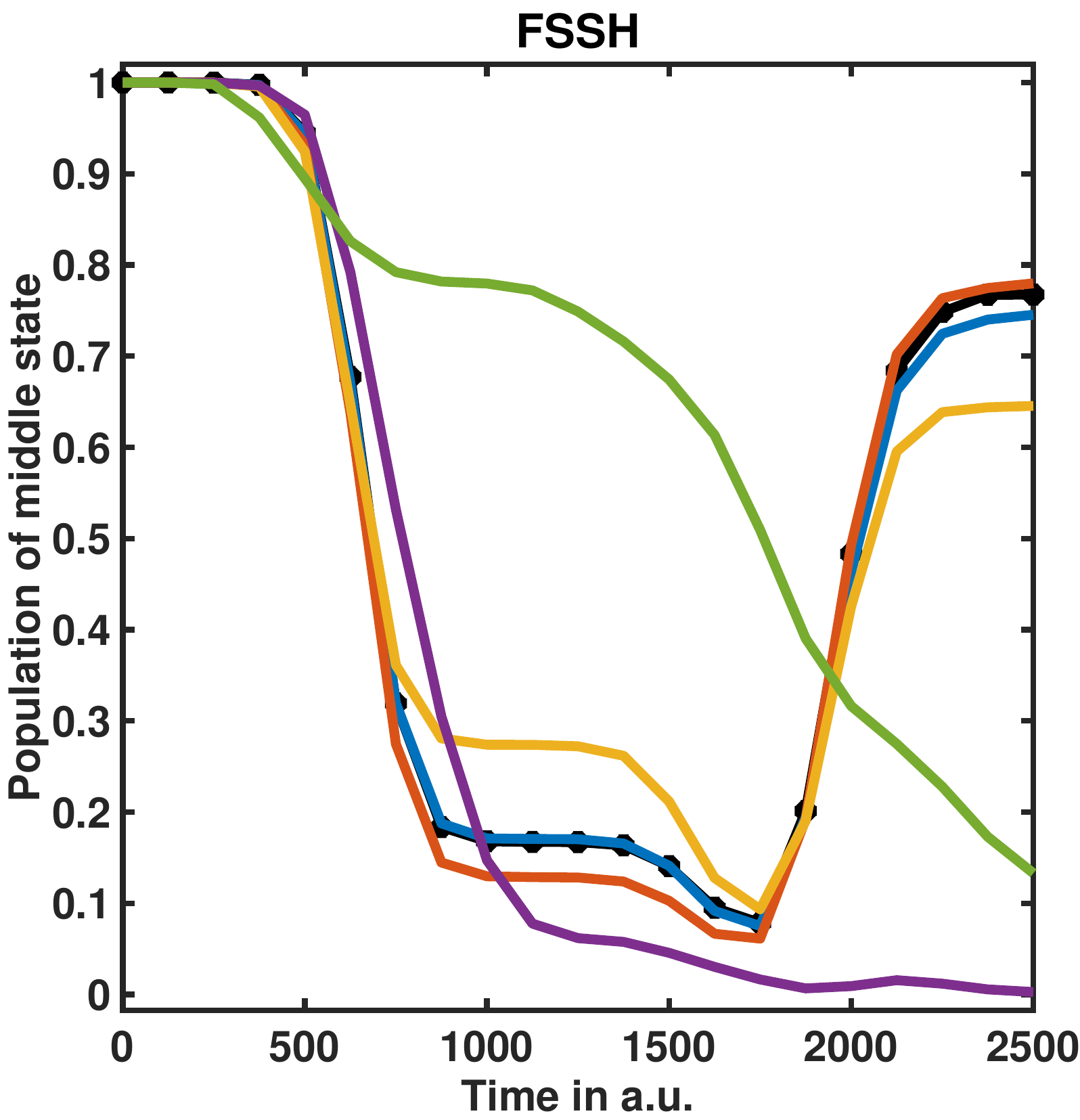}
	\includegraphics[width=0.329\linewidth] {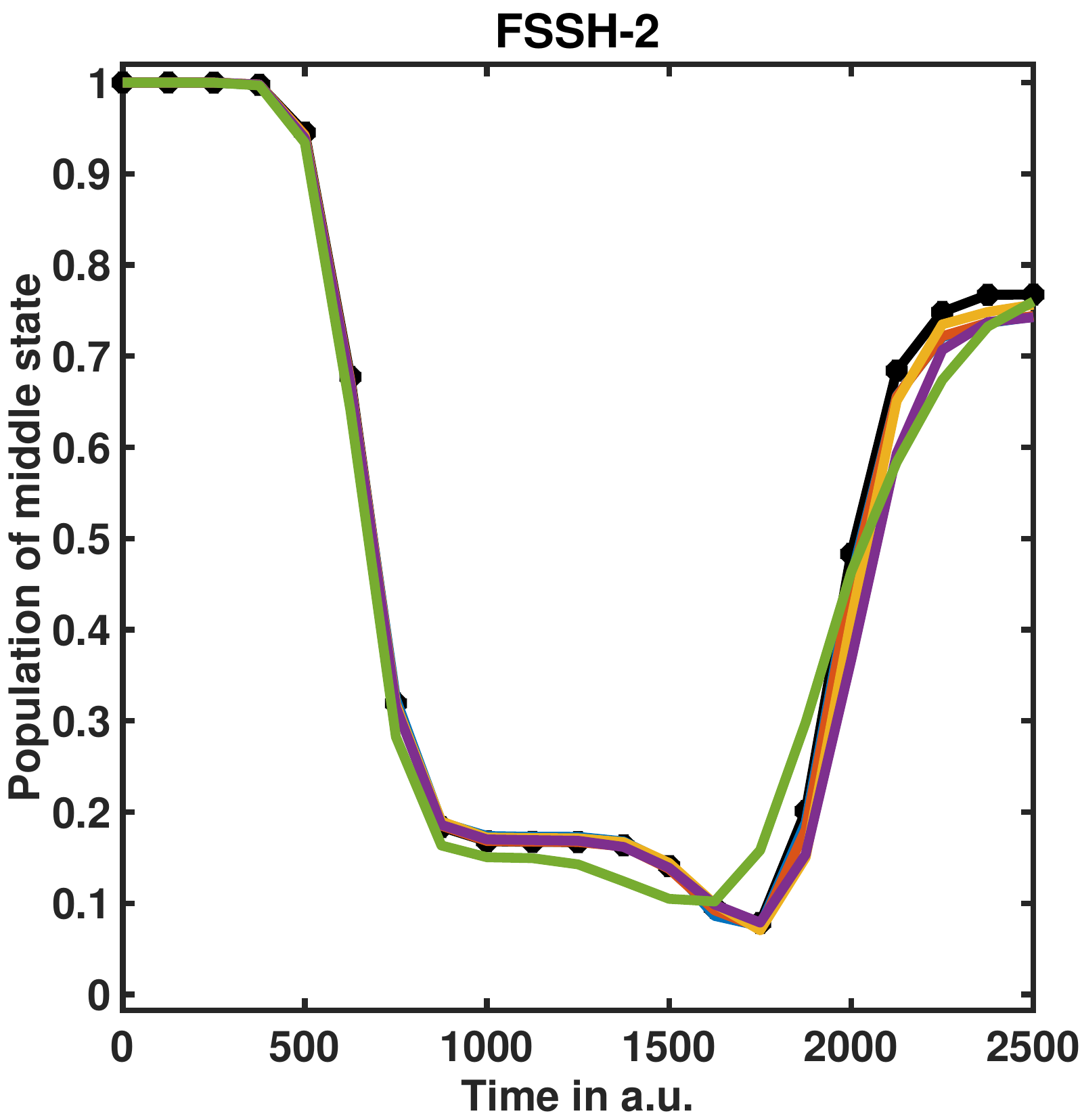}
	\includegraphics[width=0.329\linewidth] {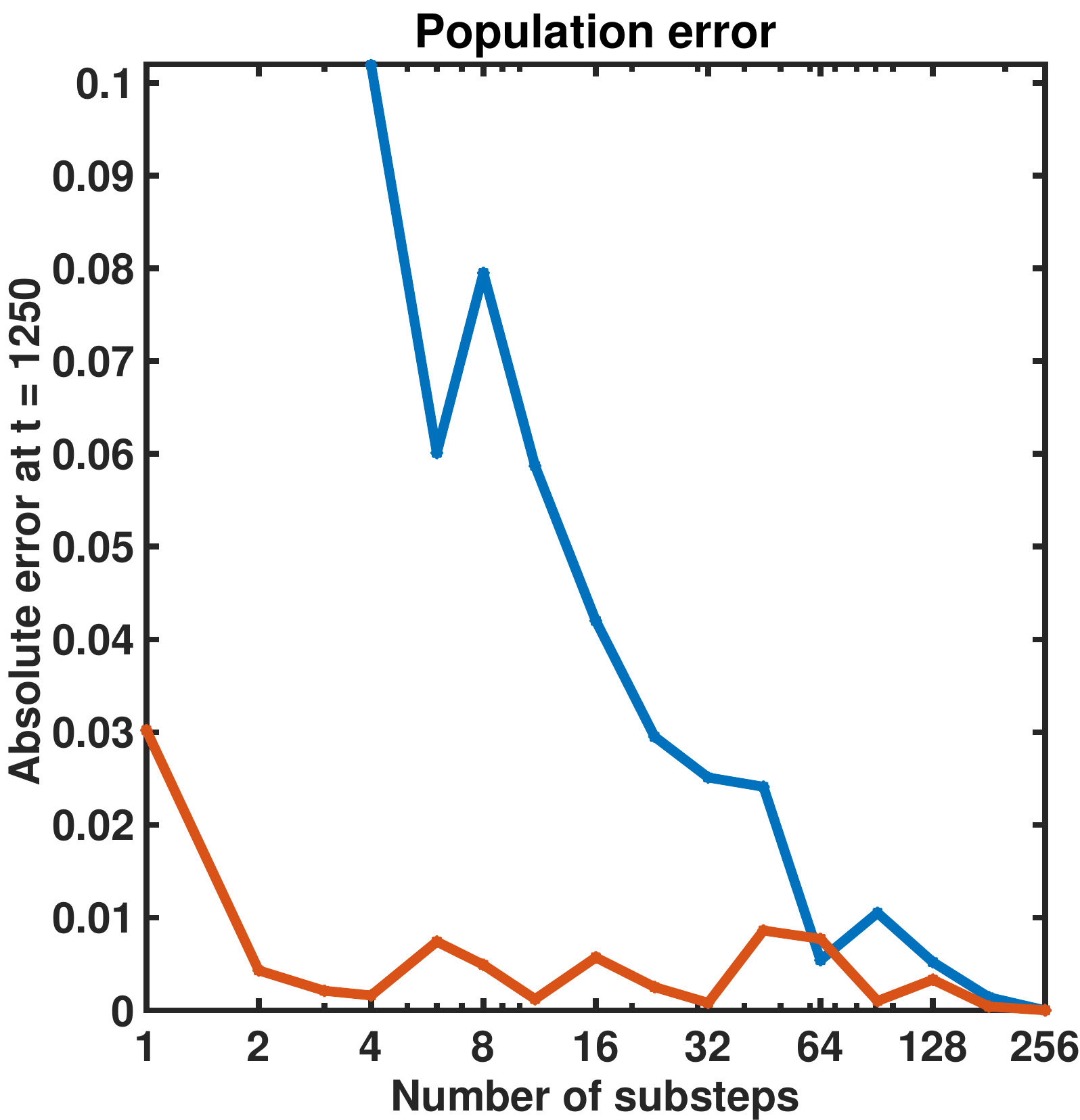}
	\caption{
		Same as Fig. \ref{fig:plotmethodconvT1} but for Model X.
	}
	\label{fig:plotmethodconvModelX}
\end{figure}

\subsection{ Comparison between FSSH and FSSH-2}
The primary aspect of our investigation involves the comparison of population transfers. 
In the case of two electronic states and under the assumption of continuous adiabatic states, FSSH-2 and FSSH exhibit a first--order approximation relationship. 
Consequently, for models featuring only two electronic states, we anticipate close accordance between FSSH and FSSH-2 results. 
The sole scenario where qualitative discrepancies might emerge is Model X because it involves three electronic states. 
Nonetheless, as shown in Tab. \ref{tbl:comparison_convergence}, across all models considered, the absolute differences in population transfers averaged over the running time generated by these methods with their maximum substep number of 256 were consistently below $0.025$.
This level of variance is acceptable for a stochastic simulation method, considering the number of trajectories involved.
Except for the 2D Well case which displayed the largest deviations, no discernible advantages were observed when these results were compared against the quantum reference solutions.

\subsection{Toward larger time step sizes}
Next, we assessed how FSSH and FSSH-2 responded to enlarged time step sizes by systematically halving the substep number, starting from 256 and descending to 1. 
Obviously, each decrease in the substep number directly cuts the total simulation time in half. 
Such a gain of efficiency has profound implications for realistic system simulations, especially when surface hopping simulations are coupled to large scale electronic structure calculations.

In Figs. \ref{fig:plotmethodconvT1} to \ref{fig:plotmethodconvModelX}, the left and center panels illustrate the population dynamics of the second adiabatic state for both the traditional FSSH and our novel FSSH-2 method, respectively.
For the standard FSSH method, results with 1, 2, and 4 substeps often significantly deviate from the converged outcomes, sometimes even presenting qualitatively incorrect results. 
In contrast, the results for our FSSH-2 method demonstrate a consistently much better performance - in most cases, the results with just a single substep per main time step align closely with the converged outcomes. 
However, the 2D LVC model is an exception. 
Initially, after the first hopping event, the FSSH-2 scheme yields nearly perfect results, but following the second hopping event, 16 substeps are required to approach the converged population results.
For fewer time steps, significant issues with the conservation of energy were found in addition to the discrepancies of the population dynamics.
To further investigate this anomaly, we re-ran the simulations for the 2D LVC model, this time rescaling the momenta along the previous momenta after each hopping event. 
While the results for the traditional FSSH method remain largely unaffected, the FSSH-2 method demonstrates enhanced accuracy: with as few as 4 substeps, convergence is reached, similarly as observed for the other models.
This observation highlights the importance of smaller time step sizes for precise handling of NACs in FSSH, particularly in multidimensional models.

To quantitatively support our observations, we assessed the maximum substep number at which the mean population deviation of the second state, calculated across all main time steps, exceeds a threshold of $0.015$, compared to the converged results. 
This threshold indicates the beginning of significant divergence. 
The traditional FSSH method typically surpassed this threshold only at relatively high substep numbers, i.~e., 4 for the Tully models and between 20 and 100 for the other models, see Tab. \ref{tbl:comparison_substeps}. 
In marked contrast, the FSSH-2 method demonstrated superior performance, exceeding the threshold already for a single substep in all models. 
The only exception was the 2D LVC model, where the FSSH-2 method surpassed this threshold only when the number of substeps was below 8. 
Hence, the FSSH-2 method showed a relative improvement over the traditional FSSH method by a factor ranging from approximately 4 to 32. 
This substantial increase in efficiency suggests a potential reduction in running time by a similar factor.

The superior stability of FSSH-2 comes to the forefront especially during the first hopping event. 
This is underscored by the graphs in the right panels of Figs. \ref{fig:plotmethodconvT1} to \ref{fig:plotmethodconvModelX} which show the population after the first passage of the crossings or intersections of the lowest two potential energy curves or surfaces.
Especially for the 2D LVC model and Model X, the FSSH-2 method offers significant advantages, allowing for time step sizes that are around 50 times greater than what FSSH requires.

Overall, these results demonstrate the superior numerical efficiency of our FSSH-2 method compared to the traditional FSSH method.

\begin{table}[t]
	\centering
	\resizebox{\textwidth}{!}{%
		\begin{tabular}{>{\centering\arraybackslash} m{2cm} >{\centering\arraybackslash} m{8cm} >{\centering\arraybackslash} m{8cm}} \toprule \toprule 
Model & FSSH & FSSH-2 \\  
 \cmidrule(lr){1-1} \cmidrule(lr){2-2} \cmidrule(lr){3-3} 
Tully 1 & 4 & 1 \\ 
Tully 2 & 4 & 1 \\ 
2D LVC & 91 & 8 \\ 
2D Well & 23 & 1 \\ 
Model X & 32 & 1 \\ 
\bottomrule \bottomrule 
\end{tabular}

	}
	\caption{
		Comparison of the FSSH and FSSH-2 methods at different substep numbers.
		The second and third columns specify the maximum substep number where the FSSH and FSSH-2 methods start to display population divergences exceeding $0.015$.
	}
	\label{tbl:comparison_substeps}
\end{table}

\section{Conclusions}

We presented the FSSH-2 scheme, an enhanced numerical time integration variant of the original FSSH method. 
The main advantage of FSSH-2 is that it is tailored to operate independently of the notorious NAC which may become (nearly) divergent at (avoided) intersections of adiabatic potential energy surfaces. 
This is known to be one of the main causes for the numerical problems hampering the original FSSH scheme, in particular the restriction to rather small time steps.
Even though our theoretical analysis indicates that the two methods do not necessarily produce identical outcomes, our simulations -- primarily based on the test suite for non--adiabatic algorithms by Nelson et al. \cite{Nelson2020b} 
-- revealed congruent results between the FSSH and FSSH-2 methods when using very small time steps.
However, the most important finding in our analysis is that the FSSH-2 scheme exhibits swifter convergence with respect to the number of time steps, which makes it approximately 1 to 2 orders of magnitude faster than conventional FSSH,
especially during the initial hopping event.

\section*{Data Availability Statement}
The MATLAB scripts used to generate the results shown in the figures and tables are openly available in the ZENODO repository at
https://doi.org/10.5281/zenodo.10497421.
While we performed the simulations using version 7.2.0 of WavePacket \cite{Schmidt2017, Schmidt2017a, Schmidt2019}, it is worth mentioning that FSSH-2 is also available in the Libra package v5.5.0 \cite{akimov_2023_10208096}.

\begin{acknowledgments}
	This work was funded by the Deutsche Forschungsgemeinschaft (DFG, German Research Foundation) through project TRR 352 – Project-ID 470903074.
	Moreover, L.~A. acknowledges support for a research stay at the WIAS in Berlin.
\end{acknowledgments}

\appendix

\section{Simulation Details} 

Throughout all of our simulations, the initial state is modeled as a Gaussian wave packet parameterized as follows
\begin{align*}
	g(q) = 
	\left(\det(W_G)\right)^{-\frac{1}{2}} 
	\left(\frac{2}{\pi}\right)^{\frac{d_{\nuc}}{4}} 
	\exp \left( - \frac{1}{4}(q-q_G)^T W_G^{-2} (q-q_G) + i p_G^T (q - q_G)\right),
\end{align*}
where \( W_G \in \mathbb{R}_{>0}^{d_{\nuc} \times d_{\nuc}} \) is a positive diagonal width matrix, determining the spread of the wave packet in each dimension. The packet is centered at position \( q_G \in \mathbb{R}^{d_{\nuc}} \), with a momentum \( p_G \in \mathbb{R}^{d_{\nuc}} \).
The initial conditions for the trajectories are derived from the Wigner transform of the Gaussian wave packet, as expressed by
\begin{align*}
	\mathcal{W}_g(q,p)
	=
	\exp \left( - 2 (q-q_G)^T W_G^{-2} (q-q_G) - \frac{1}{2} (p-p_G)^T W_G^{2} (p-p_G) \right)
	,
\end{align*}
implying that the sampling follows a normal distribution in $q$ and $p$.

\subsection{Tully 1}
A single crossing model consisting of two electronic states
with diabatic potential matrix given by \cite{Tully1990}
\begin{align*}
	V^{\dia}(q)
	&=
	\begin{pmatrix} 
		\sign q A \left(1-\exp(- \sign q B q)\right) & C \exp(-D q^2) \\ 
		C \exp(-D q^2) & - \sign q A \left(1-\exp(- \sign q B q)\right)
	\end{pmatrix} 
\end{align*}
with $A = 0.01$, 
$B = 1.6$,
$C = 0.005$,
$D = 1.0$,
and mass $M = 2000$.
The initial Gaussian wave packet has the parameters $W_G = 0.75$, $q_G = -6$, $p_G = 15$.
Initially, the electronic system occupies the lower adiabatic state.
The main time step size is $200$ a.u. for a total of $9$ main time steps.
The main time step size is $100$ a.u. for a total of $10$ main time steps. 
The grid for the quantum solution consisted of $2^8$ equally spaced points in the range $[-10,10]$.

\subsection{Tully 2}
A double crossing model consisting of two electronic states
with diabatic potential matrix given by 
\cite{Tully1990}
\begin{align*}
	V^{\dia}(q)
	&=
	\begin{pmatrix} 
		0 & C \exp(-D q^2) \\ 
		C \exp(-D q^2) & - A \exp(- B q^2) + E_0
	\end{pmatrix} 
\end{align*}
with $A = 0.1$, 
$B = 0.28$,
$C = 0.015$,
$D = 0.06$,
$E = 0.05$,
and mass $M = 2000$.
The initial Gaussian wave packet has the parameters $W_G = 0.75$, $q_G = -7$, $p_G = 30$.
Initially, the electronic system occupies the lower adiabatic state.
The main time step size is $100$ a.u. for a total of $10$ main time steps. 
The grid for the quantum solution consisted of $2^9$ equally spaced points in the range $[-12,12]$.

\subsection{2D LVC}
A conical intersection model consisting of two electronic states
with diabatic potential matrix given by 
\cite{Gherib2016}
\begin{align*}
	V^{\dia}(q_1,q_2)
	&=
	\frac{1}{2}
	\begin{pmatrix} 
		\omega_1^2 (q_1+\frac{a}{2})^2 + \omega_2^2 q_2^2 + \sigma & 2 c q_2 \\ 
		2 c q_2 & \omega_1^2 (q_1-\frac{a}{2})^2 + \omega_2^2 q_2^2 - \sigma
	\end{pmatrix} 
\end{align*}
with the parameters used for bis(methylene) adamantyl cation: 
$\omega_1 = 7.743 \cdot 10^{-3}$,
$\omega_2 = 6.680 \cdot 10^{-3}$,
$a  = 31.05$,
$c  = 8.092 \cdot 10^{-5}$,
$\sigma  = 0$,
and mass $M = 1$.
The initial Gaussian wave packet has the parameters $W_G = \diag (2 \omega_1, 2 \omega_2)^{- \frac12}$, $q_G = (a/2,0)^T$, $p_G = (0,0)^T$.
Initially, the electronic system occupies the upper adiabatic state.
The main time step size is $100$ a.u. for a total of $11$ main time steps. 
The grid for the quantum solution consisted of $2^8 \times 2^7$ equally spaced points in the range $[-80,80] \times [-40,40]$.

\subsection{2D Well}
A 2D extension of the dual avoided crossing model (Tully 2, see above) consisting of two electronic states 
with diabatic potential matrix given by
\cite{Shenvi2011}
\begin{align*}
	V^{\dia}(q_1,q_2)
	&=
	\begin{pmatrix} 
		\scriptstyle -E_0 
		& \scriptstyle C \exp \left( - D \left( 0.25 (q_1+q_2)^2 + 0.75 (q_1-q_2)^2 \right) \right) 
		\\ 
		\scriptstyle C \exp \left( - D \left( 0.25 (q_1+q_2)^2 + 0.75 (q_1-q_2)^2 \right) \right) 
		&\scriptstyle -A \exp \left( - B \left( 0.75 (q_1+q_2)^2 + 0.25 (q_1-q_2)^2 \right) \right)
	\end{pmatrix} 
\end{align*}
with $A   = 0.15$, $B   = 0.14$, $C   = 0.015$, $D   = 0.06$, $E_0 = 0.05$, and mass $M = 2000$.
The initial Gaussian wave packet has the parameters $W_G = \frac{1}{\sqrt{2}} \id_{2 \times 2}$, $q_G = (-8,0)^T$, $p_G = (20,0)^T$.
Initially, the electronic system occupies the lower adiabatic state.
The main time step size is $100$ a.u. for a total of $13$ main time steps. 
The grid for the quantum solution consisted of $2^8 \times 2^7$ equally spaced points in the range $[-15,15] \times [-10,10]$.

\subsection{Model X}
A triple avoided crossing model consisting of three electronic states
with diabatic potential matrix given by
\cite{Subotnik2011}

\begin{align*}
	\scriptstyle 
	V^{\dia}(q)
	&=
	\begin{pmatrix} 
		\scriptstyle A \left( \tanh \left( B q \right) + \tanh \left( B (q+7) \right) \right)
		& \scriptstyle C \exp \left( - q^2 \right) 
		& \scriptstyle C \exp \left( - (q+7)^2 \right) 
		\\
		\scriptstyle C \exp \left( - q^2 \right)
		& \scriptstyle -A \left( \tanh \left( B q \right) + \tanh \left( B (q-7) \right) \right)
		& \scriptstyle C \exp \left( - (q-7)^2 \right)
		\\
		\scriptstyle C \exp \left( - (q+7)^2 \right)
		& \scriptstyle C \exp \left( - (q-7)^2 \right)
		& \scriptstyle -A \left( \tanh \left( B (q+7) \right) - \tanh \left( B (q-7) \right) \right)
	\end{pmatrix} 
\end{align*}
with $A = 0.03$, 
$B =1.6$,
$C = 0.005$,
and mass $M = 2000$.
The initial Gaussian wave packet has the parameters $W_G = 0.75$, $q_G = -12$, $p_G = 15$.
Initially, the electronic system occupies the central adiabatic state.
The main time step size is $125$ a.u. for a total of $20$ main time steps. 
The grid for the quantum solution consisted of $2^9$ equally spaced points in the range $[-15,25]$.

\bibliography{references}
\end{document}